\documentclass{article}
\usepackage[utf8x]{inputenc}
\usepackage{spconf,amsmath,graphicx}

\graphicspath{{figures/}}

\usepackage{cite}
\usepackage{amsmath,amssymb,amsfonts}
\usepackage{algorithmic}
\usepackage{graphicx}
\usepackage{multirow}
\usepackage{tikz}
\usepackage{hyperref}
\usepackage{xcolor}
\usepackage{lipsum}
\usepackage{caption}
\usepackage{siunitx}
\usepackage{commath}
\usepackage{subfig}
\usepackage{pgfplots}
\pgfplotsset{compat=1.17}
\usepackage{mathrsfs}
\usetikzlibrary{arrows}
\usetikzlibrary{positioning}
\usetikzlibrary{3d}
\usepackage{textcomp}
\usepackage{xcolor}
\def\BibTeX{{\rm B\kern-.05em{\sc i\kern-.025em b}\kern-.08em
    T\kern-.1667em\lower.7ex\hbox{E}\kern-.125emX}}
 
\usepackage{hyperref}
\hypersetup{
	colorlinks=true,
	linkcolor=blue,
	filecolor=magenta,      
	urlcolor=cyan,
	pdftitle={Flexible-Rate Learned Hierarchical Bi-directional Video Compression with Flow Refinement},
	pdfauthor={M. A. Yılmaz; O. Keleş; H. Güven; M. Tekalp},
	pdfstartview=Fit,
	pdffitwindow=True,
}

\begin{document}
\definecolor{col_w}{rgb}{0.870588,0.796078,0.776470}
\definecolor{col_wn}{rgb}{0.996078,0.847059,0.364706}
\definecolor{col_out}{rgb}{1,0.5,0}
\definecolor{col_bc}{rgb}{0,0.5,1}
\definecolor{col_conv}{rgb}{0,1,1}
\definecolor{col_out}{rgb}{1,0.5,0}
\title{Flexible-Rate Learned Hierarchical Bi-directional Video Compression with Motion Refinement and Frame-Level Bit Allocation}

\name{Eren Çetin, M. Akın Yılmaz, A. Murat Tekalp
\thanks{This work was supported in part by TUBITAK 2247-A Award No. 120C156 and KUIS AI Center funded by Turkish Is Bank. A. M. Tekalp also acknowledges support from Turkish Academy of Sciences (TUBA).}}

\address{Dept. of Electrical \& Electronics Engineering and KUIS AI Center, Koç University, Istanbul, Turkey}

\maketitle

\begin{abstract}
This paper presents improvements and novel additions to our recent work on end-to-end optimized hierarchical bi-directional video compression~\cite{lhbdc} to further advance the~state-of-the-art in learned video compression. As an improvement, we combine motion estimation and prediction modules and compress refined residual motion vectors for improved rate-distortion performance. As novel addition, we adapted the~gain unit proposed for image compression to flexible-rate video compression in two ways: first, the~gain unit enables a single encoder model to operate at multiple rate-distortion operating points; second, we exploit the~gain unit to control bit allocation among intra-coded vs. bi-directionally coded frames by fine tuning corresponding models for truly flexible-rate learned video coding. Experimental results demonstrate that we obtain state-of-the-art rate-distortion performance exceeding those of all prior art in learned video coding.
\end{abstract}

\vspace{2pt}

\section{Introduction}
\label{intro}
Following the pioneering work~\cite{balle2017endtoend} on  variational autoencoder-based end-to-end optimized image compression, significant rate-distortion (R-D) performance improvements have been obtained to achieve state-of-the-art results in image compression by using better entropy modeling~\cite{balle2018variational, minnen_joint, variational_low, cheng2020image, minnen2020channelwise}.

End-to-end learned video compression frameworks employ learned image compression models for intra-frame coding, and typically replace the functional blocks used in P~and/or B-picture mode of standards-based video codecs, such as motion estimation, motion compensation, motion vector compression and residual compression, with their learned counterparts. Sub-networks corresponding to these blocks are jointly trained using a single R-D loss. One of the first low-latency video coding models is DVC~\cite{lu2019dvc}, which performs optical flow-based backwarping for motion compensation and employs a post-processing network to alleviate motion compensation errors. Scale-space flow (SSF)~\cite{agustsson_scale} introduces the scale channel, which applies blurring to regions where flow estimates are unreliable. In contrast to DVC and SSF, which use a single past reference frame, RLVC~\cite{rlvc} and ELF-VC~\cite{elfvc} utilize a recurrent architecture to exploit long-term temporal correlations. A detailed review on end-to-end video compression frameworks can be found in~\cite{video_review}.

This paper addresses the combination of bi-directional hierarchical video coding and flexible rate coding, which are less studied in the learned compression community. The~literature on these topics is reviewed in Section~\ref{related}. The proposed method is introduced in Section~\ref{BVC-FR}. Our main contributions are to improve the compression efficiency of our previous work~\cite{lhbdc} by motion refinement and residual motion coding and also employ gain unit for flexible rate coding using a single end-to-end optimized model. Experimental results are presented in Section~\ref{eval}. Finally, Section~\ref{conc} concludes the paper.


\section{Related work and Contributions}
\label{related} \vspace{-4pt}

\subsection{Bi-directional Video Compression}
Hierarchical Learned Video Compression (HLVC)~\cite{hlvc} is a bi-directional codec that employs three hierarchical quality layers and a recurrent enhancement network. LHBDC~\cite{lhbdc}, proposed concurrently, performs bi-directional motion compensation and obtains superior R-D performance compared to HLVC. Racape et al.~\cite{spie_bidir} later used a similar approach using conditional convolutions for video compression in YCrCb~420 space. 
Ladune et al.~\cite{ladune2021conditional} implemented a conditional coder to process I, P and B frames using a single network, which ignores unavailable elements. In this paper, we extend our previous work~\cite{lhbdc} for improved performance and added flexible rate functionality as detailed in Section~\ref{ss:contrib}. \vspace{-17pt}

\subsection{Flexible Rate Image/Video Compression}
Most learned image/video compression frameworks require training separate models for different rate-distortion points, which increases training cost and memory requirement. 
As a promising solution to this problem, AG-VAE~\cite{Cui_2021_CVPR} proposed learned gain units to scale the latent representation in the channel dimension with gain vectors prior to quantization for flexible-rate image compression. After learning the gain vectors for different bitrates, exponential interpolation is proposed to achieve continuous rate adaptation during inference. ELF-VC~\cite{elfvc} propose a different flexible-rate framework allowing a single model to cover a large and dense range of bitrates, at a negligible increase in computation and parameter count. We adopt the former approach in this paper and extended it for the video compression setting.
\vspace{-6pt}

\subsection{Contributions}
\label{ss:contrib}
Our previous work~\cite{lhbdc} presented an end-to-end optimized (for a particular R-D operating point) B-frame encoder, called LHBDC, which relied on a pre-trained optical flow estimation model. This paper proposes a truly end-to-end optimized flexible-rate video encoder with the following contributions: \\
1) The proposed new B-frame encoder model does not rely on a pretrained motion estimation network. Instead, we propose an autoencoder model for direct flow prediction and a flow refinement approach for flow field compression. \\
2) We adapt the gain unit \cite{Cui_2021_CVPR} proposed for flexible rate image coding to achieve continuous R-D curve in video coding. \\
3) We fine-tune the bitrate allocation between intra-coded and bi-directionally coded frames to achieve a frame-level bitrate allocation scheme that yields superior results in the R-D sense. To the best of our knowledge, this is the first paper to exploit the gain unit for frame-level bitrate allocation.


\section{Flexible-rate Bi-directional Video Compression with Motion Refinement}
\label{BVC-FR}
\begin{figure*}[ht]
\centering
	\includegraphics[width=0.95\textwidth]{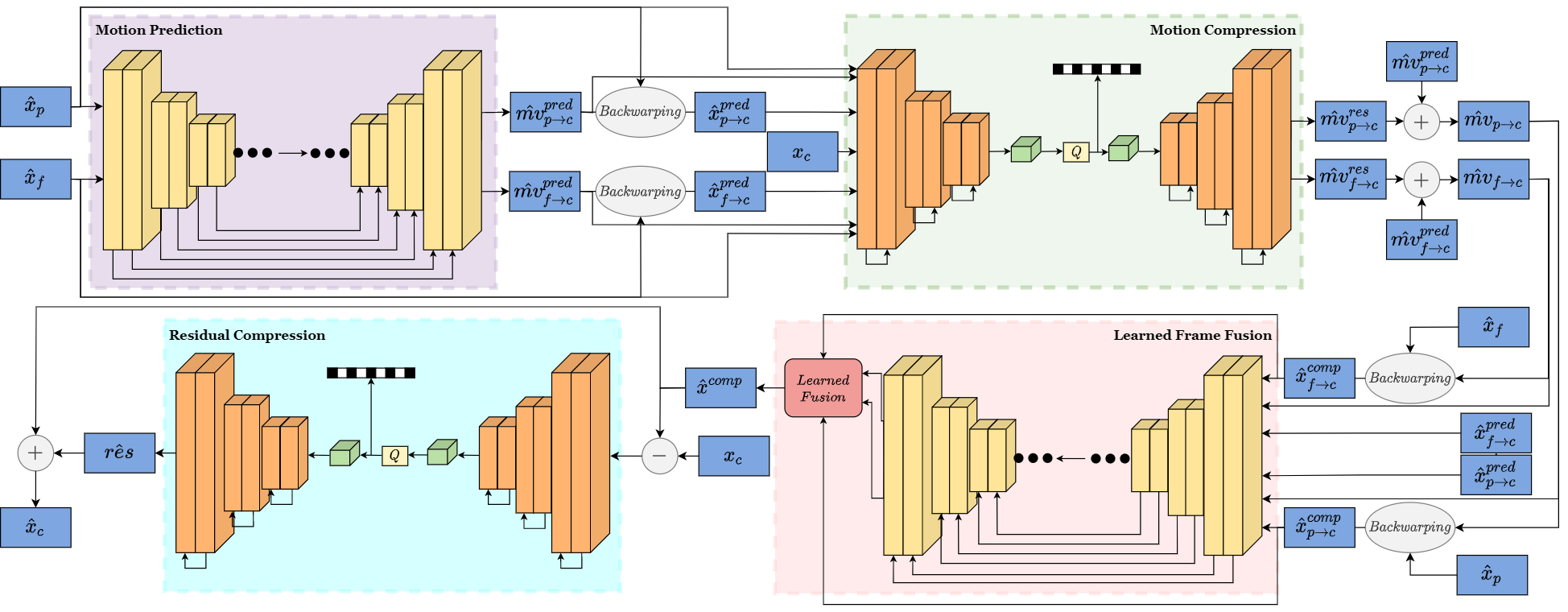} \vspace{-6pt} \\
\caption{Overview of the proposed learned bi-directional video encoder network. The motion prediction and learned frame fusion modules have 5 and 4 layers of stacked convolution blocks (yellow layers), respectively, with leaky ReLUs as activation functions. The motion compression and residual compression modules utilize a variational autoencoder with residual blocks (orange layers) and gain units (green blocks) to learn quantization parameters while achieving low entropy latent representations.}
\label{fig:framework}
\end{figure*}

We propose a network composed of four modules,  depicted in Figure~\ref{fig:framework}, for bi-directional B-frame compression, which does not depend on any pre-trained component, given a past and a future reference frame. We introduce motion prediction in Section~\ref{mot_pred}, motion compression in Section~\ref{mot_comp}, learned frame fusion in Section~\ref{learn_fusion}, residual compression in Section~\ref{res_comp}, and gain unit in Section~\ref{gain_unit}.

The first frame of each group-of-pictures (GoP) is compressed as keyframe using the model proposed by Minnen et al.~\cite{minnen_joint} without the context model while rest of the frames are compressed bi-directionally using the proposed network. 
We fine-tune the I-frame model and the proposed B-frame model jointly to achieve optimal frame-level bitrate allocation in the~R-D sense benefiting from the gain unit.

\subsection{Motion Prediction}
\label{mot_pred}
We utilize the U-Net architecture with 5 layers, depicted in Figure~\ref{fig:framework} (upper left), to predict bi-directional motion vectors. This module takes the past and future decoded frames, $\hat{x}_{p}$ and $\hat{x}_{f}$, respectively, and yields predicted motion vectors, $\hat{mv}^{pred}_{c\rightarrow p}$ and $\hat{mv}^{pred}_{c\rightarrow f}$, from the~current frame $x_c$ towards the past and future decoded frames. 
We then compute two predictions for the current frame by first
bilinear backwarping the past decoded frame, $\hat{x}_{p}$, using $\hat{mv}^{pred}_{c\rightarrow p}$, and second, backwarping the~future decoded frame, $\hat{x}_{f}$, using 
$\hat{mv}^{pred}_{c\rightarrow f}$ to generate the predicted current frames, $\hat{x}^{pred}_{p\rightarrow c}$ and $\hat{x}^{pred}_{f\rightarrow c}$, given by
\begin{align}
    \label{eq:final_motvec}
    \hat{x}^{pred}_{p\rightarrow c}=W_b(\hat{x}_p,\hat{mv}^{pred}_{c\rightarrow p})\\
    \hat{x}^{pred}_{f\rightarrow c}=W_b(\hat{x}_f,\hat{mv}^{pred}_{c\rightarrow f})
\end{align}
where $W_b()$ denotes the backwarping operator.

\subsection{Motion Refinement and Compression}
\label{mot_comp}
The motion compression module, depicted in Figure~\ref{fig:framework} (upper right), learns bi-directional residual motion vectors between the current frame and two predicted current frames, $\hat{x}^{pred}_{p\rightarrow c}$ from the past and $\hat{x}^{pred}_{f\rightarrow c}$ from the future reference, and encodes them. We learn the residual motion vectors as the output of an autoencoder, similar to that used in the scale-space flow model~\cite{agustsson_scale}. This module employs residual blocks with $N=128$ filters each. Novelties compared to~\cite{agustsson_scale} are: i) we adapt the model in~\cite{agustsson_scale} to bi-directional motion-compensation, ii) we estimate residual/refined flow vectors unlike~\cite{agustsson_scale}, which estimates flow field in one shot. iii) We do not use scale channel as occlusions are taken care by the frame fusion module. In order to compress the residual motion vectors, we employ a hyperprior network similar to~\cite{minnen_joint} to learn entropy parameters. 

The inputs to this module are the predicted motion vectors, $\hat{mv}^{pred}_{c\rightarrow p}$ and $\hat{mv}^{pred}_{c\rightarrow f}$, the predicted current frames, $\hat{x}^{pred}_{p\rightarrow c}$ and $\hat{x}^{pred}_{f\rightarrow c}$, the decoded reference frames, $\hat{x}_{p}$ and $\hat{x}_{f}$, and the~ground-truth current frame, $x_{c}$. The decoder part of the~autoencoder generates bi-directional motion residual vectors, $\hat{mv}^{res}_{c\rightarrow p}$ and $\hat{mv}^{res}_{c\rightarrow f}$ as outputs. These residual motion vectors are added to the predicted motion vectors to compute the final bi-directional motion vectors given by
\begin{align}
    \hat{mv}_{c\rightarrow p}=
    \hat{mv}^{pred}_{c\rightarrow p}+\hat{mv}^{res}_{c\rightarrow p} \label{eq:eq3}\\
    \hat{mv}_{c\rightarrow f}=\hat{mv}^{pred}_{c\rightarrow f}+\hat{mv}^{res}_{c\rightarrow f} \label{eq:eq4}
\end{align}

\subsection{Motion Compensation by Learned Frame Fusion}
\label{learn_fusion}
The refined bi-directional motion vectors (\ref{eq:eq3})-(\ref{eq:eq4}) are used to backwarp the past and future reference frames to obtain higher quality forward and backward motion compensated frames, compared to one-shot motion, which are input to a learned mask generator network similar to one used in~\cite{lhbdc}. 

The learned frame fusion module, which is a U-Net with 4 layers, generates two masks, $m_1$ and $m_2$, with the same dimensions as video frames. The inputs to the module are the motion compensated frames, $\hat{x}^{comp}_{p\rightarrow c}$ and $\hat{x}^{comp}_{f\rightarrow c}$, the refined motion vectors, $\hat{mv}_{c\rightarrow p}$ and $\hat{mv}_{c\rightarrow f}$, and the reference frames, $\hat{x}_p$ and $\hat{x}_f$. Given the masks, $m_1$ and $m_2$, a single motion compensated frame is computed as
\begin{align}
\begin{split}
    \label{eq:masking}
    \hat{x}^{comp}=\;
    &\dfrac{m_1}{m_1+m_2}\odot W_b(\hat{x}_p, \hat{mv}_{c\rightarrow p})\;+\\
    &\dfrac{m_2}{m_1+m_2}\odot W_b(\hat{x}_f, \hat{mv}_{c\rightarrow f})
\end{split}
\end{align}

\subsection{Motion-compensated Residual Frame Compression}
\label{res_comp}
The motion-compensated  residual frame is computed by subtracting the ground truth current frame, $x_c$ from the motion compensated frame, $\hat{x}^{comp}$ and is compressed by an autoencoder network that is similar to the motion compression module with $N=128$ filters at each layer. 
After we decode the encoded latent representation of the residual frame, we compute the current frame by adding the motion compensated frame, $\hat{x}^{comp}$ and the reconstructed residual frame, $\hat{res}$.

\subsection{Gain Unit for Flexible-Rate Coding}
\label{gain_unit}
The gained network architecture~\cite{Cui_2021_CVPR} enables a single model to operate at multiple R-D points by learning scaling parameters per latent channel for discrete rate-distortion levels during training. The learned scaling parameters for each discrete level are stored in vectors called ``gain" and ``inverse gain" units. As the latent representations for the motion residuals and frame residuals both have $N=128$ channels, we scale the latent representations with two separate vectors of length $L=128$ in the channel dimension. This way, we effectively change the quantization bin sizes before rounding the~latent representation to the nearest integer at inference time. The~inverse quantization operation is performed on the quantized latent representation using separate learned vectors of length $L=128$ at the decoder.
We also employ gain and inverse gain units in the respective hyper-prior networks, which learn and compress the entropy parameters in a separate bottleneck. Using the gain units, we can control the bitrate allocated to motion and frame residuals separately. In addition, we can also control the bitrates allocated to intra-coded and  bi-directionally coded individual frames.

\section{Evaluation}
\label{eval}
The proposed network is implemented using the PyTorch~\cite{pytorch} framework and is employed to compress  bi-directional predicted frames. The convolution layers in compression modules have $C=128$ channels, while the number of channels in motion prediction and frame fusion modules are 64 and 32, respectively. The keyframes are compressed with the~pretrained model~\cite{minnen_joint} from the CompressAI library~\cite{compressai}.
The details of the training process is explained in~Section 4.1. Experimental results and comparison with the state of the art are presented in Section 4.2. 

\subsection{Training Details}
The training is performed on the Vimeo-90k dataset~\cite{vimeo}. The~overall network shown in Figure~\ref{fig:framework} is trained jointly using a single R-D loss with 4 discrete R-D levels
\begin{align}
\label{eq:loss}
    L= \sum\limits_{n=1}^{4}\left(\lambda_n \cdot D + R\right)
\end{align}
where \textit{$\lambda_i=0.0067, 0.025, 0.048, 0.093$} are the
Lagrange multipliers. The gain and inverse gain vectors for both motion and frame residual compression modules as well as for their hyper-prior networks are paired with the respective Lagrange multipliers. Hence, we have 4 gain and 4 inverse gain vectors for each compression module as well as its hyper-prior network. 

For training, we crop random patches of $256 \times 256$ from frames of size $256 \times 448$ for data augmentation. The mini-batch size is set to 4 and the model is optimized with respect to mean squared error using Adam optimizer~\cite{adam} for 1M iterations. To prevent overshooting, we clip the norm of the gradients so that the maximum norm value is 1. The initial learning rate is set to 1e-4 and a learning rate scheduler is deployed so that the learning rate is halved if there is no improvement for 100K iterations.

\begin{figure}[t]
\centering
	\includegraphics[scale=0.33]{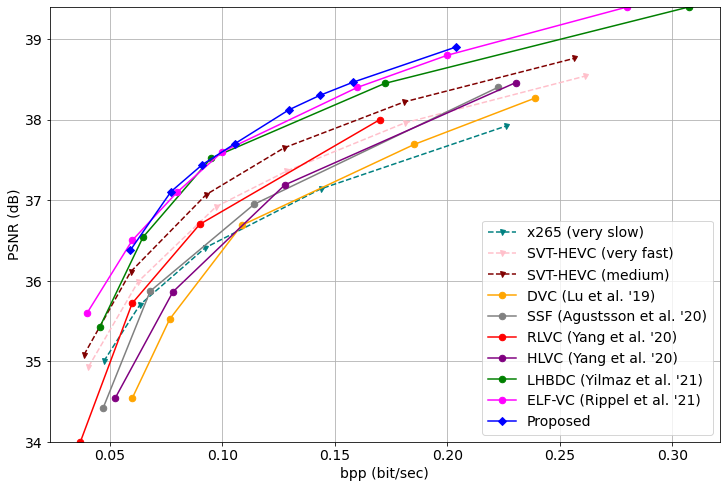} \vspace{-14pt} \\
\caption{Comparison of R-D performance of the proposed model with prior works in terms of PSNR.}
\label{fig:psnr_curve}
\end{figure}

\begin{figure}[t]
\centering
	\includegraphics[scale=0.25]{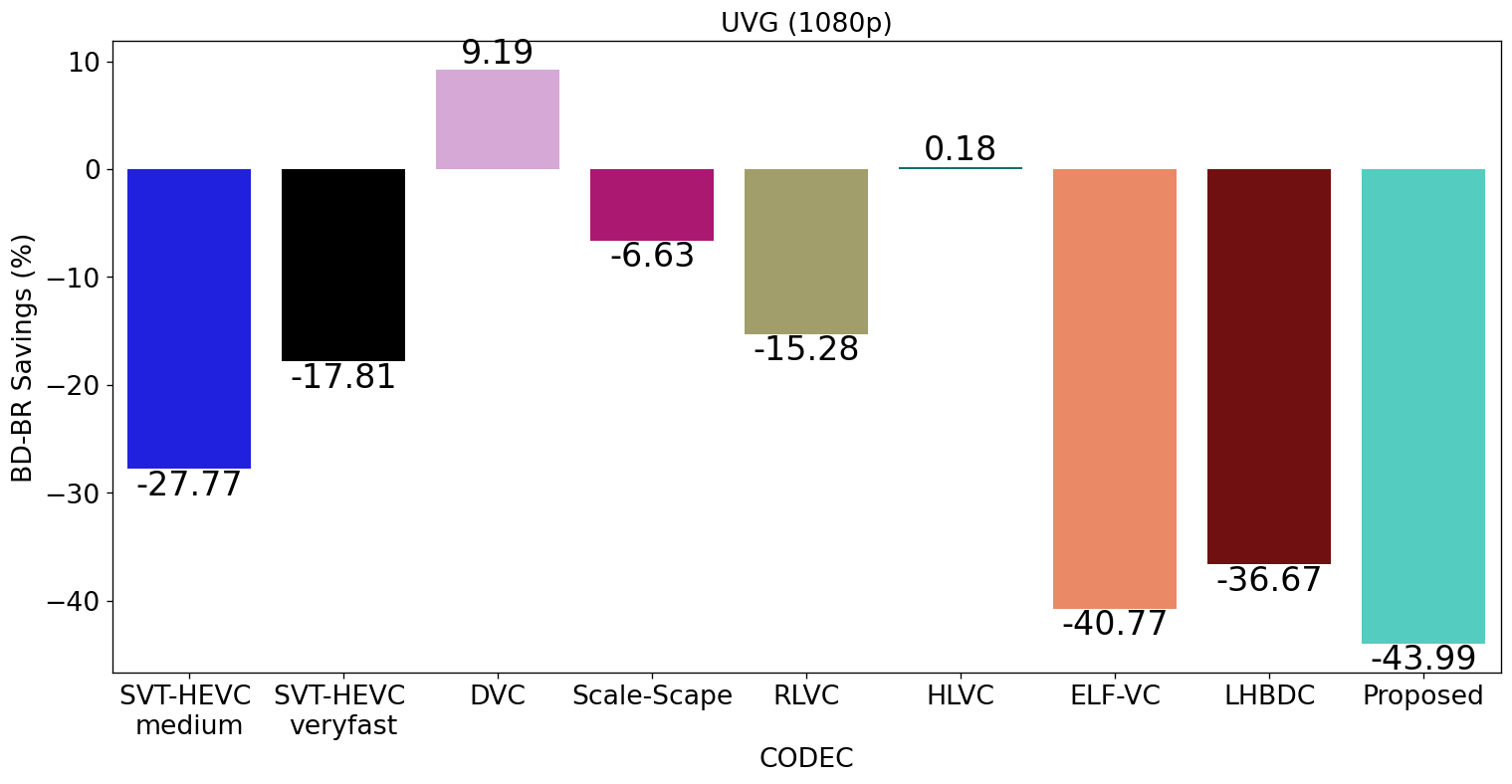} \vspace{-6pt} \\
\caption{Average percent BD-BR improvements (RGB bpp) for the proposed model and other video codecs vs. the anchor x265 encoder (veryslow preset) on the UVG dataset.}
\label{fig:rate_savings}
\end{figure}

\subsection{Experimental Results}
The proposed model with GoP size 16 is evaluated on the UVG dataset~\cite{uvg}. The rate-distortion (R-D) performance of our model, i.e., the peak signal-to-noise ratio (PSNR) vs. bits per RGB pixel, is compared with those of other leading learned video compression models and traditional H.265/HEVC video codecs in Figure~\ref{fig:psnr_curve}.

R-D points for the proposed model are achieved by performing exponential interpolation on the adjacent gain vectors. For example, to determine an R-D point between the gain vectors, $v_1$ and $v_2$, we interpolate an intermediate gain vector using 
\begin{align}
\label{eq:exp_interp}
v_l=v_1^l\odot~v_2^{(1-l)}
\end{align}
where $0<l<1$ corresponds to the interpolation coefficient. As our model can achieve arbitrarily many R-D points using this process with varying $l$, $v_1$ and $v_2$, we aimed finding the optimal bitrate allocation of hierarchical levels that yields the optimal R-D curve. Our experimental results revealed that the superior R-D curve was achieved by lowering quality levels for bi-directional coded frames by intervals of 0.33 so that $l_i=l_{i-1}-0.33$ where $l_i$ is the interpolation coefficient of the hierarchical level $i$. 

When we compare the resulting R-D performance in terms of PSNR, our proposed network achieves superior results at higher bitrates while it attains similar performance at lower bitrate regions compared to ELF-VC~\cite{elfvc}. On the other hand, it indubitably outperforms other learned and traditional video codecs that are present in Figure~\ref{fig:psnr_curve}.

BD-BR improvements~\cite{bdrate} over other leading learned and standards-based video codecs are presented in Figure~\ref{fig:rate_savings}. The~proposed model achieves \%~43.99 BD-BR improvement over the anchor x265 encoder  (veryslow preset) while other codecs achieve lower gains. 

Finally, ablation studies reveal the significance of the proposed motion prediction and learned frame fusion improvements on the R-D performance. The motion prediction module reduces the entropy of motion vectors to be encoded while the learned frame fusion alleviates occlusion artifacts. The improvements due to both modules are presented in Table~\ref{table:ablation}.

\begin{table}
\centering
\caption{Ablation study on motion prediction and learned frame fusion modules on the UVG dataset.}  \vspace{-4pt}
\begin{tabular}{clrl} 
\hline
Property                                                                        & \multicolumn{1}{c}{Option} & \multicolumn{1}{l}{\begin{tabular}[c]{@{}l@{}}BD-BR \\Increase\end{tabular}} & \multicolumn{1}{c}{\begin{tabular}[c]{@{}c@{}}Total \# \\of Param
\label{table:ablation}
\end{tabular}}  \\ 
\hline
\multirow{2}{*}{Motion Predictor}                                               & ~\underline{Yes}               & 0\%~                                                                           & ~ 35M                                                                            \\
                                                                                & ~No                        & \textcolor{red}{37\%~}                                                         & \textcolor[rgb]{0,0.502,0}{~ 23M (-34+)}                                         \\ 
\hline
\multirow{2}{*}{\begin{tabular}[c]{@{}c@{}}Learned Frame \\Fusion\end{tabular}} & ~\underline{Yes}               & 0\%~                                                                           & ~ 35M                                                                            \\
                                                                                & ~No                        & \textcolor{red}{23\%~}                                                         & \textcolor[rgb]{0,0.502,0}{~ 32M (-9\%)}                                         \\
\hline
\end{tabular}
\end{table}
\section{Conclusion}
\label{conc}
\vspace{-3pt}

We propose an end-to-end optimized flexible-rate hierarchical bi-directional video compression network that is trained once to operate at multiple bitrates. The resulting single learned model yields superior R-D performance in terms of PSNR vs. RGB bits/sec from very low to high bitrates. 

The proposed network demonstrates improvements by employing motion prediction and learned frame fusion while achieving arbitrarily many R-D points using the gain and inverse gain units that scale latent representations in the channel dimension effectively. Compared to other learned and traditional video codecs, the proposed network also achieves the best BD-BR reduction on the UVG dataset.

To the best of our knowledge, this is the first study on frame-level bitrate allocation for learned codecs. We aim to further investigate optimal rate allocation among intra-coded and bi-directional coded frames jointly as future work.


\clearpage
\bibliography{references}
\bibliographystyle{IEEEtran}
\end{document}